\documentclass[a4paper]{jacow}
\usepackage{bm}

\makeatletter%
    \ifboolexpr{bool{xetex}}
     {\renewcommand{\Gin@extensions}{.pdf,%
                        .png,.jpg,.bmp,.pict,.tif,.psd,.mac,.sga,.tga,.gif,%
                        .eps,.ps,%
                        }}{}
\makeatother

\ifboolexpr{bool{xetex} or bool{luatex}} 
 {}                                      
 {\usepackage[utf8]{inputenc}}           

\usepackage[USenglish]{babel}

\begin{document}

\title{Beam energy measurement at the European XFEL with high-performance trajectory
    fitting}

\author{L.~Fröhlich\thanks{lars.froehlich@desy.de}, Deutsches Elektronen-Synchrotron DESY,
    Hamburg, Germany}

\maketitle

\begin{abstract}
    This contribution presents the method used for beam energy monitoring at the European
    XFEL. It relies on high-performance trajectory fitting of beam position monitor data
    against an online model of the accelerator lattice. Unlike most other methods, it can
    be applied to virtually any section of the lattice with sufficient dispersion without
    the need for manual adaptation. Our implementation is fast enough to measure the
    energy of many thousands of bunches per second in realtime on conventional x86
    hardware shared with other processes. A short outline of the technique is presented
    and illustrated with data from the accelerator.
\end{abstract}

\section{Introduction}

Good knowledge of the beam energy is important for practically all particle accelerators.
Not only does a deviation from the lattice design energy translate into focusing and
trajectory errors, but usually the accelerator-driven experiments themselves need to know
the energy with sufficient precision: Particle physics experiments require the
center-of-mass energy, in light sources the wavelength of the emitted radiation depends
directly on the particle energy, and so on.

A multitude of techniques for measuring the beam energy have been developed over the
years. Before describing the method used at the European XFEL, we will have a quick look
at other well-known techniques that rely solely on basic beam dynamics and on beam
position measurements.

In a synchrotron, the stored beam follows a closed orbit and therefore the beam energy
can be estimated relatively well by requiring the total deflection angle of all dipoles
to be 360°:
\begin{equation*}
    \frac{q}{p} \int\!B\,\mathrm{d}s = 2\pi,
\end{equation*}
where $q$ is the charge of the particle, $p$ the beam momentum, and $B$ the magnetic flux
density. Obviously, better knowledge of the integrated magnetic field
$\int\!B\,\mathrm{d}s$ leads to better energy precision. Therefore, in some cases,
an especially well calibrated spectrometer magnet is inserted into the lattice to
facilitate precision measurements (e.g., Ref.~\cite{Ass2005}). The deflection angle of this
single dipole is then measured by beam position monitors (BPMs) surrounding it.

The latter technique is also typical for single-pass energy measurements in linear
accelerators (linacs) or transfer lines, sometimes with variations in beam position
diagnostics (e.g., Ref.~\cite{Ken1989}). The type of lattice also varies --- often, beamlines
are not specifically designed as spectrometers, but have a high dispersion that can
simply be exploited for energy measurements as well. In single-pass free-electron lasers
(FELs), magnetic four-dipole chicanes serve as bunch compressors, but at the same time
offer an ideal location for measuring the beam energy. In such machines, angle and
offset of the incoming beam are rarely negligible and need to be considered. While
analytical expressions for such corrections can be constructed using linear optics
(e.g., Ref.~\cite{See1999}), they are always specific to a given lattice.

This paper presents a method for the measurement of the beam energy that, like most
standard methods, relies on beam position monitors and on a good knowledge of all
optical elements in the section under examination. Its main advantage is that it can be
applied to virtually \emph{any} section of an accelerator lattice without needing to be
adapted specifically.

\section{Trajectory Fitting Algorithm}

\begin{figure}
    \centering
    \includegraphics[width=\linewidth]{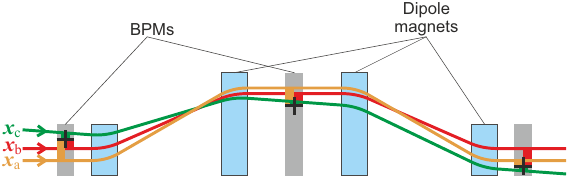}
    \caption{Illustration of the trajectory fitting algorithm for a four-dipole chicane.
             The black crosses in the BPMs show the measured beam position. The algorithm
             tracks multiple particles with varying start conditions $\bm{x}$ through the
             lattice and finds the start condition with the minimum quadratic difference
             between measured and simulated beam positions. In the illustration, the green
             trajectory with start condition $\bm{x}_c$ yields the best reproduction of
             the measurement.}
    \label{fig:trajectory_fitting_in_chicane}
\end{figure}

The method is based on an iterative \emph{trajectory fitting algorithm} that is applied to
the beamline segment (lattice) of interest. Figure~\ref{fig:trajectory_fitting_in_chicane}
illustrates the algorithm: First, the beam positions $x^\mathrm{meas}_i, y^\mathrm{meas}_i$
are acquired from all $N$ BPMs in the lattice (black crosses). An optics simulation code
with knowledge of the current state of magnets and other components then tracks a test
particle through the lattice (yellow curve). The particle is launched with a certain set
of initial coordinates\footnote{We use the common conventions for the coordinates in the
curvilinear coordinate system of the beam: $x$, $y$ are transverse offsets relative to the
design trajectory, $x'$, $y'$ the corresponding angles of motion, $s$ the longitudinal
offset, and $\delta = \Delta E / E$ the relative energy deviation of the particle.}
$\bm{x} = \left(x, x'\!, y, y'\!, s, \delta\right)$ and its transverse position
$x^\mathrm{sim}_i, y^\mathrm{sim}_i$ is stored at each BPM along the way.

The quadratic difference of the simulated and measured positions,
\begin{equation*}
    \xi = \sum_{i = 1}^N \left(x^\mathrm{meas}_i - x^\mathrm{sim}_i\right)^2 +
                         \left(y^\mathrm{meas}_i - y^\mathrm{sim}_i\right)^2 \,\text{,}
\end{equation*}
is a measure of how well the simulation reproduces the measured trajectory, lower values
indicating better agreement. A multi-dimensional minimization method with $\xi(\bm{x})$ as
its objective function can therefore be used to find the launch condition $\bm{x}$ that
approximates the observed trajectory most closely. In other words, the algorithm keeps
launching test particles under variation of the initial coordinates (red and green curves)
until it finds the best match to the BPM readings. The most interesting result of this
trajectory fit for our application is the $\delta$ component of the coordinate vector,
which translates directly into the deviation from the design energy $E$ of the
magnets: $\Delta E = E \delta$.

\section{Implementation Details}

The algorithm is implemented in a server for the DOOCS\,\cite{DOOCSwww, Gry1996}
control system. It reads beam position data from shared memory via the data acquisition
(DAQ\,\cite{Wil2008a}) system of the accelerator. The server and the associated particle
tracking and numerical libraries are written in C++. They can deliver energy measurements
in realtime for several hundred different electron bunches in different locations along
the accelerator and along the bunch train at the 10~Hz repetition rate of the European
XFEL\,\cite{Agh07}, sharing standard x86-64 server hardware with multiple other processes.

A Nelder-Mead simplex algorithm\,\cite{Nel1965} is used for the minimization problem.
Depending on the lattice configuration, only a subset of the launch coordinates is
actually varied: In the case of the bunch compressors, varying $(y, y'\!, \delta)$ is
sufficient because their geometry lies entirely in the $y$--$z$ plane and there are no
sources of $x$--$y$ coupling. The typical number of needed iterations is in the range of
$\sim$40--110. For the fits of consecutive bunches in the train, this number can be
reduced substantially by using the results from the previous bunch as initial values.

A separate server handles the setup and monitoring of the chicane magnets. This software
ensures that the origin of the vertical energy BPM\,\cite{Lor2018} readout is zero for a
particle entering the chicane exactly on axis and with the design energy of the chicane.
It does so by calculating the vertical displacement of the trajectory from the deflection
angle and subtracting it from the absolute vertical position measured by the BPM system.

\section{Measurement Uncertainties}

Like most standard methods, the trajectory fitting technique relies on good BPMs and on a
decent knowledge of all optical elements in the section under examination. As the
objective function of the fit is the quadratic sum of the differences between measured and
simulated beam positions, there are many error sources, most notably:
\begin{minipage}{\linewidth}
\vspace{0.5cm}
\begin{Itemize}
    \item BPM offsets (on the measurement side),
    \item Quadrupole and higher multipole offsets,
    \item Magnetic field uncertainties, especially those of corrector magnets which are
        usually not cycled (standardized).
\end{Itemize}
\end{minipage}

All of these error sources contribute directly to the \emph{systematic error} of the
resulting energy measurement. Obviously, the systematic error depends on the specific
lattice section. At the European XFEL, we estimate it as $\sim$0.5\,\% for most measurement
locations. With a lattice section specifically designed for energy measurement and with
carefully aligned components, much better precision should be possible. The
\emph{statistical error} is essentially determined by the noise of the individual BPMs. By
the nature of the fit, it acts like an averaging filter between the various beam positions
entering into it. For the high-energy part of the XFEL accelerator, the relative
statistical error of the energy measurement is on the order of $3\cdot10^{-5}$.

\section{Operational Experience}

Since its initial development phase, the beam energy measurement server has been in
operation at the European XFEL for $\sim$8~years. Its feature set and the amount of users
have grown considerably over time. Currently the server measures the beam energy
simultaneously in 17 separate lattice segments. The machine accelerates trains of up to
2700~bunches per RF pulse. Nine representative bunches out of this train are processed in
realtime (i.e.\ at the full repetition rate of 10\,Hz). Fitting of the full train is only
done on demand. Depending on the number of bunches and on the overall system load, it can
take up to a few seconds for the full train results to become available.

\begin{figure}[!h]
    \centering
    \includegraphics[width=\linewidth]{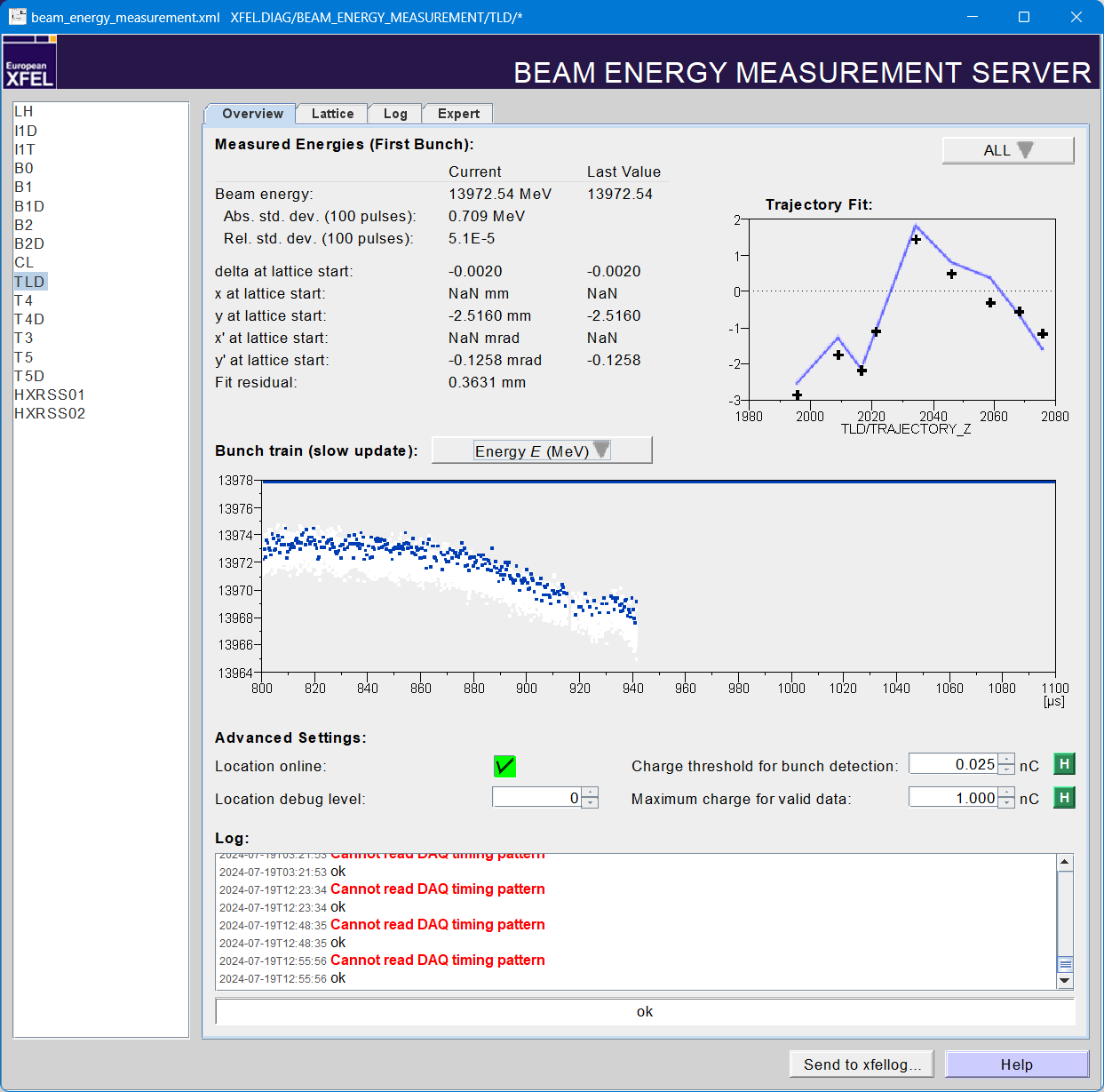}
    \caption{Screenshot of a user panel for the beam energy measurement server.}
    \label{fig:screenshot}
\end{figure}

Figure~\ref{fig:screenshot} shows a screenshot of a user panel for the server. The
trajectory fit in the top right corner is for the first bunch of the train in the highly
dispersive European XFEL linac dump line. The energy along the bunch train is shown in the
center of the window, and a small droop towards the end of the train is visible.

The algorithm and its implementation are very robust. We routinely operate several energy
feedbacks on it; most~\cite{Kam2019} use the fast output for the representative bunches,
but one stabilizes the energy profile over the train at the end of the linac based on the
slower train data.

The same energy measurement server runs at the FLASH free-electron laser~\cite{Vog13}
where it serves 10 different accelerator sections and processes 10 representative bunches
out of the train in realtime.

\section{Summary}

Energy measurement via the trajectory fitting technique is just one of the many
applications of a continuously updating online model of the machine lattice. For some of
the sections in which we use it, analytic solutions with much lower computational demands
could be devised. Still, even on commercial off-the-shelf hardware the fitting method
is fast enough to provide realtime measurements. Its main advantage, however, is its
generality. It can be applied to virtually \emph{any} section of an accelerator lattice
without needing to be adapted specifically. Even nonlinear and special lattice elements
are handled seamlessly as long as they are present in the online model, and adding new
measurement sections is as easy as identifying a suitable start and end component.

\section{Acknowledgements}

The author wishes to thank Bolko Beutner and Johann Zemella for many helpful discussions
on the algorithm and on its application for the European XFEL and FLASH, respectively.
The work was inspired by a similar project of our esteemed late colleague Vladimir
Balandin.

The author acknowledges support from DESY (Hamburg, Germany), a member of the Helmholtz
Association HGF. This work was supported by operation funds of the European XFEL.

\end{document}